\documentclass[%
 reprint,
 amsmath,amssymb,
 aps,
 prl
]{revtex4-2}
\usepackage{graphicx}
\usepackage{amsmath}
\usepackage{amssymb}
\usepackage{bm}
\usepackage{physics}
\usepackage{comment}
\usepackage{textcomp}
\usepackage{times}
\usepackage{color}
\usepackage{soul}

\begin{document}
\preprint{APS/123-QED}

\title{Demonstration of geometric diabatic control of quantum states}

\author{Kento Sasaki}
\email{kento.sasaki@phys.s.u-tokyo.ac.jp}
\affiliation{Department of Physics, The University of Tokyo, Bunkyo-ku, Tokyo 113-0033, Japan}
\author{Yuki Nakamura}
\affiliation{Department of Physics, The University of Tokyo, Bunkyo-ku, Tokyo 113-0033, Japan}
\author{Tokuyuki Teraji}
\affiliation{National Institute for Materials Science, Tsukuba, Ibaraki 305-0044, Japan}
\author{Takashi Oka}
\affiliation{The Institute for Solid State Physics, The University of Tokyo, Kashiwa, Chiba 277-8581, Japan}
\author{Kensuke Kobayashi}
\email{kensuke@phys.s.u-tokyo.ac.jp}
\affiliation{Department of Physics, The University of Tokyo, Bunkyo-ku, Tokyo 113-0033, Japan}
\affiliation{Institute for Physics of Intelligence, The University of Tokyo, Bunkyo-ku, Tokyo 113-0033, Japan}
\affiliation{Trans-scale Quantum Science Institute, The University of Tokyo, Bunkyo-ku, Tokyo 113-0033, Japan}

\date{\today}

\begin{abstract}
Geometric effects can play a pivotal role in streamlining quantum manipulation.
We demonstrate a geometric diabatic control, that is, perfect tunneling between spin states in a diamond by a quadratic sweep of a driving field.
The field sweep speed for the perfect tunneling is determined by the geometric amplitude factor and can be tuned arbitrarily.
Our results are obtained by testing a quadratic version of Berry's twisted Landau-Zener model.
This geometric tuning is robust over a wide parameter range.
Our work provides a basis for quantum control in various systems, including condensed matter physics, quantum computation, and nuclear magnetic resonance.
\end{abstract}


\maketitle 

\begin{figure}
\begin{center}
\includegraphics{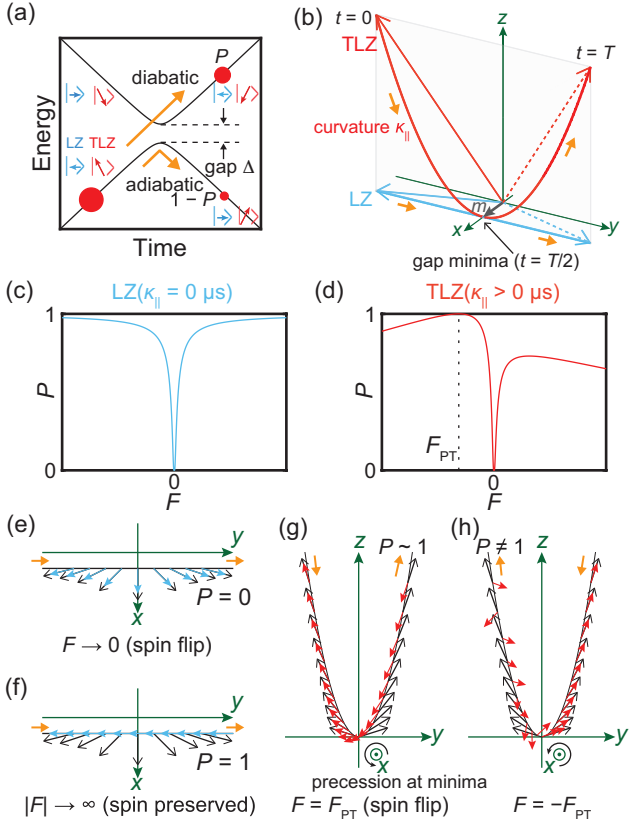}
\caption{
Comparison of the Landau-Zener (LZ) transition and the twisted Landau-Zener (TLZ) transition.
(a) Tunneling at level anti-crossing.
(b) Sweeping of the driving field.
The fields at $t=0$ and $t=T$ for the LZ (TLZ) model are indicated by solid and dashed blue (red) arrows, respectively.
Predicted (c) LZ transition probability and (d) TLZ transition probability are plotted as a function of the speed $F$.
Dynamics of the field (black arrow) and spin (blue arrow) in the LZ model are plotted in the (e) adiabatic and (f) diabatic limits [see (b) and (c)].
Dynamics of the field (black arrow) and spin (red arrow) in the TLZ model at (g) $F=F_\text{PT}$ and (h) $F=-F_\text{PT}$ [see (b) and (d)].
The solid black line indicates the field amplitude in the $xy$ plane in (e) and (f) and in the $yz$ plane in (g) and (h).
The origin of each arrow corresponds to the field amplitude at each instant.
\label{fig1}
}\end{center}
\end{figure}

Tunneling is an exotic yet ubiquitous quantum phenomenon. 
To control quantum states, a common strategy known as adiabatic control avoids it by moving a large energy barrier slowly.
Another ubiquitous feature of quantum physics is geometric effects~\cite{Wilczek1989}. 
A well-known example is the geometric phase~\cite{berry1984diabolical} that a particle acquires during an adiabatic motion. 
However, geometric effects are not restricted by adiabaticity. 
Even during diabatic tunneling events, geometric effects take place and lead to grave consequences in the dynamics.

The simplest system that demonstrates the marriage of tunneling and geometric effects is the twisted Landau-Zener (TLZ) model introduced by Berry, which describes a particle in two quantum states driven by an external field~\cite{berry1990geometric}. 
In the original untwisted Landau-Zener (LZ) model~\cite{landau1932theorie,zener1932non,stuckelberg1932theorie,majorana1932atomi}, 
when two energy levels change in time, 
quantum tunneling across an energy gap $\Delta$ occurs depending on the speed of the change [Fig.~\ref{fig1}(a)] \cite{shevchenko2010landau,ivakhnenko2022landau}.
The tunneling probability $P$ depends on the sweep speed $F$ [Fig.~\ref{fig1}(c)]; $P=0$ in the adiabatic limit ($F\rightarrow0$), while $P$ is unity in the diabatic limit ($\abs{F}\rightarrow\infty$).  
Such speed-dependent tunneling has been demonstrated in various systems~\cite{oliver1999hanbury,petta2010coherent,huang2011landau,zhou2014observation,gasparinetti2011geometric,zhang2014realization,wang2016experimental}.
In the TLZ model, the driving field has a ``twist" and the adiabatic to diabatic transition is geometrically modulated~\cite{zhao2008adiabatic,xu2018quantized}.
Recently, the importance of the geometric effects was recognized not only in equilibrium~\cite{xiao2010berry} but also in nonequilibrium~\cite{ma2021topology,basov2017towards,orenstein2021topology,de2021colloquium}.
The TLZ model, which possesses a new nonequilibrium tuning knob on top of the LZ model, should be widely applied to materials engineering~\cite{takayoshi2021nonadiabatic,kitamura2020nonreciprocal} and quantum controls~\cite{gaitan2003temporal,gaitan2004controlling,ran2007high,li2011robust}.
Despite a few experiments~\cite{zwanziger1991measuring,bouwmeestert1996observation,bouwmeester1996observation,zwanziger2003non}, the opportunity to utilize such geometric tuning for quantum control has long been overlooked, and its robustness remains unexplored.

Here, using an electron spin in a diamond, we realize and test an ideal TLZ model with a quadratic twist~\cite{takayoshi2021nonadiabatic} that manifests perfect tunneling and nonreciprocity over a wide range of gap and twist parameters.
We measure the tunneling probabilities with high precision and obtain an average of 95.5~\% under the condition where perfect tunneling occurs.
The condition of perfect tunneling can be smoothly tuned by adjusting the curvature of the quadratic sweep.
These geometrical effects are robust beyond the framework of the existing theory~\cite{takayoshi2021nonadiabatic}.
This geometric diabatic control is ubiquitous and can be applied to various quantum systems.

As a geometric diabatic control, we aim to realize perfect tunneling ($P=1$) and change the state at the same time. 
The Hamiltonian for the TLZ model in the natural units is defined as~\cite{takayoshi2021nonadiabatic},
\begin{align}
\hat{H} &= \bm{b}\cdot \bm{\hat{\sigma}} = m \hat{\sigma}_x + \nu q \hat{\sigma}_y + \frac{1}{2}\kappa_{\parallel} \nu^2q^2 \hat{\sigma}_z,
\label{eq1}
\end{align}
where $\hat{\sigma}_j$ ($j=x$, $y$, and $z$) is the Pauli operator, and $\bm{b} = (b_x, b_y, b_z) \equiv (m, \nu q, 1/2 k_\parallel \nu^2 q^2)$ is a driving field.
We change the parameter $q$ in time as $q=-F(t-T/2)$ between time $t=0$ and $t=T$ with a dimensionless sweep speed $F$.
This is a quadratic version of the original TLZ model~\cite{berry1990geometric}; $\Delta=2m$ is the gap, and $2\nu$ ($>0$) is the energy slope.
Figure~\ref{fig1}(b) depicts the initial and final fields as a red solid arrow ($t=0$) and a red dotted arrow ($t=T$), respectively.
The $b_z$ component, which depends quadratically on time, induces a ``twist" of the field.
This twist appears in the trajectory of the field [the solid red line in Fig.~\ref{fig1}(b)] and its strength is determined by the geodesic curvature $\kappa_{\parallel}$.
Situations in which the spin and driving field are always kept parallel or antiparallel are adiabatic; situations that deviate from this are diabatic.
The diabatic geometric effect is captured by the 
geometric amplitude factor~\cite{berry1990geometric} (also known as the quantum geometric potential~\cite{zhao2008adiabatic} or shift vector~\cite{kitamura2020nonreciprocal})
$R_{12}(q)=-A_{11}(q)+A_{22}(q)+\partial_q{\rm arg}A_{12}(q)$, where the Berry connection is defined by  $A_{nl}(q)=\bra{n(q)}i\partial_q\ket{l(q)}$ using the instantaneous eigenstate $\ket{n(q)}$ satisfying $\hat{H}(q)\ket{n(q)}=E_n(q)\ket{n(q)}$.  
The tunneling probability $P$ from $\ket{1}$ to $\ket{2}$ is given by~\cite{takayoshi2021nonadiabatic},
\begin{equation}
P \approx 
 \exp\left[-\frac{\pi}{4\nu \abs{F}}\left(\Delta+\frac{FR_{12}(0)}{2}\right)^2\right],
\label{eq2}
\end{equation}
where $R_{12}(0)=\nu \kappa_{\parallel}$ holds in the present model. 
Equation~(\ref{eq2}), referred to as ``TLZ formula" in this work, is derived using a twisting coordinate transformation~\cite{takayoshi2021nonadiabatic}
and it recovers the LZ formula when $\kappa_{\parallel}=0$ [Fig.~\ref{fig1}(c)]. 
We stress that the TLZ formula is approximate in contrast to the LZ formula which is asymptotically exact. 
Figure~\ref{fig1}(d) shows the behavior of the transition described by the TLZ formula when $\kappa_{\parallel}>0$.
The $P$ is nonreciprocal to the sign reversal of the speed $F$ corresponding to the field sweep direction~\cite{kitamura2020nonreciprocal}.
In  Eq.~(\ref{eq2}), the gap $\Delta$ in the LZ model is effectively shifted to $\Delta+\frac{FR_{12}(0)}{2}$ by the geometric amplitude factor~\cite{zhao2008adiabatic,takayoshi2021nonadiabatic}.
In particular, when the speed is,
\begin{equation}
F_\text{PT} = -2\Delta/R_{12}(0)\qquad 
\label{eqpt}
\end{equation}
the effective gap closes and the tunneling probability saturates $P \approx 1$.
We call this behavior ``perfect tunneling (PT)"~\cite{takayoshi2021nonadiabatic}, and the speed at which $P$ is maximized is referred to as the ``PT condition".
In contrast to the LZ case, the quantum state changes during the diabatic transition from the initial state
$\ket{1(q=FT/2)}$ to the final state $\ket{2(q=-FT/2)}$, and thus allows us to realize geometric diabatic control of the quantum states. 
Our main purpose is to extensively test the behaviors predicted by the TLZ formula [Eq.~(\ref{eq2})].

\begin{figure}
\begin{center}
\includegraphics{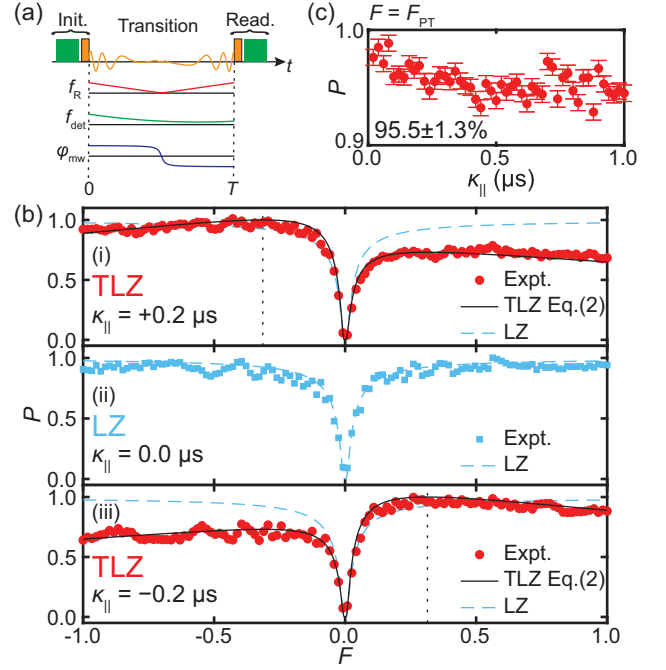}
\caption{
Demonstration of the TLZ transition at $m=0.5$~MHz.
(a) Measurement sequence.
Laser and microwave pulses are used for initialization and readout of the NV center.
(b) Dependence of tunneling probability $P$ on speed $F$.
The squares and circles indicate experimental results, black solid lines indicate the TLZ formula [Eq.(\ref{eq2})], and vertical black dotted lines indicate $F=F_\text{PT}$.
The blue dashed lines indicate the LZ formula [TLZ with $R_{12}(0)=0$].
(c) Tunneling probability at $F=F_\text{PT}$ in the range of $\kappa_{\parallel} = 0 \text{--} 1~\mu$s.
The error bars indicate 65~\% confidence intervals estimated from the shot noise of the PL measurement.
\label{fig2}
}\end{center}
\end{figure}

We realize the TLZ transition with an electron spin of a single nitrogen-vacancy (NV) center in a diamond~\cite{fuchs2011quantum,huang2011landau,zhou2014observation}.
We use the NV center's $m_S=0$ and $-1$ states as a two-level system and manipulate it with microwave pulses.
In a suitable rotating frame (see Supplemental Material~\cite{sm}), the Hamiltonian is expressed as ($\hat{S}_i$ denotes the $S=\frac{1}{2}$ spin operators)
\begin{align}
\hat{H}_\text{r} &= f_\text{R} \qty[ \cos(\phi_\text{mw}) \hat{S}_x
-\sin(\phi_\text{mw}) \hat{S}_y ] + \frac{d (f_\text{det}t)}{dt} \hat{S}_z,
\label{eq3}
\end{align}
where $f_\text{R}$ is the Rabi frequency corresponding to the microwave field amplitude, $\phi_\text{mw}$ is the microwave phase, and $f_\text{det}$ is the detuning between the resonance frequency and the microwave frequency.
We generate a microwave pulse satisfying $f_\text{R} = \sqrt{b_x^2  + b_y^2}$, $\phi_\text{mw} = -\arctan(b_y/b_x)$, and $f_\text{det} = \int_0^{t} b_z(t') dt'$, so that Eq.~(\ref{eq3}) reproduces the driving field $\bm{b}$ in the TLZ Hamiltonian [Eq.~(\ref{eq1})].
This conversion to the $S=\frac{1}{2}$ system in MKS units corresponds to making the following changes to each parameter; $m\rightarrow\pi m$, $\nu \rightarrow\pi \nu$, and $\kappa_{\parallel}\rightarrow \kappa_{\parallel}/\pi$ (see~\cite{sm}).
We adjust the sweep duration $T$ considering the coherence time and available microwave parameter ranges.
Figure~\ref{fig2}(a) shows the measurement sequence.
We use green laser pulses and photoluminescence (PL) measurements for spin initialization and readout.
We prepare the initial and final states using rectangular microwave pulses after and before the laser pulse to match the instantaneous field direction with the projection direction.
The obtained PL intensity is precisely converted to a tunneling probability using reference PL intensities of the $m_S=0$ and $=-1$ states~\cite{misonou2020construction}.

We show our experimental results obtained when the gap parameter is fixed as $m=0.5$~MHz.
Without loss of generality, we investigate the probability $P$ [Eq.~(\ref{eq2})] by selecting the energy slope $\nu$ to $(10~\mathrm{MHz})^2$ and adjusting only the dimensionless speed $F$.
First, we set $\kappa_{\parallel}=0~\mu$s to address the conventional LZ model.
The blue circles in Fig.~\ref{fig2}(b ii) show the experimental result.
The lower the speed ($F\rightarrow0$), the lower the transition probability $P$; the behavior is symmetric between positive and negative speeds.
It agrees well with the LZ formula (black solid line) and proves that our system reproduces the LZ model with high accuracy (for more details see~\cite{sm}).

We then address the TLZ transition when $\kappa_{\parallel}=+0.2~\mu$s shown in Fig.~\ref{fig2}(b i).
The experimental result (red circles) is asymmetric in $F\to -F$ and becomes higher for $F<0$ than for $F>0$.
The $P$ reaches maxima in the vicinity of the predicted PT condition ($F=F_\text{PT}$) indicated by the vertical dashed line.
Specifically, as shown in Fig.~\ref{fig2}(c), we find $P=95.5\pm1.3~\%$, on average, in a range of $\kappa_{\parallel}=0.0 \text{--} 1.0~\mu$s.
Figure~\ref{fig2}(b iii) shows the results when $\kappa_{\parallel}=-0.2~\mu$s.
Compared to the $\kappa_{\parallel}=+0.2~\mu$s case [Fig.~\ref{fig2}(b i)], it shows totally inverted behavior to the speed $F$.
These behaviors are qualitatively different from the LZ transition (blue dashed line) and well reproduced by the TLZ formula without any adjustable parameters (black solid line).
These are our central results, proving that the tunneling probability is successfully modulated by the geodesic curvature $\kappa_{\parallel}$ of the driving field, resulting in perfect tunneling and nonreciprocity.
The fact that perfect tunneling, which has only been possible in the extremely fast speed limits of the LZ model, is achieved even at finite speeds is essentially different in the long history of the LZ physics.

Here we give an intuitive picture of the perfect tunneling phenomenon.
Figure~\ref{fig1}(g) shows the driving field (black arrow) and spin (red arrow) dynamics at $F=F_\text{PT}$.
The quadratic sweep produces adiabatic dynamics in the initial stage ($t\sim 0$) and diabatic dynamics near the gap minima ($t\sim T/2$).
Near the gap minima, the $x$ component of the driving field $\bm{b}$, i.e., the gap itself ($b_x=m$), causes spin precession and rotates the spin around the $x$ axis.
When the PT condition is fulfilled, this rotation of the spin is synchronized with the counterclockwise twist of the field (also around the $x$ axis) and the transition to the excited state is achieved smoothly.
Thus a spin flipping is realized [Fig.~\ref{fig1}(g)].
When the sweep direction is reversed ($F=-F_\text{PT}$), as shown in Fig.~\ref{fig1}(h), the clockwise field twist cannot synchronize with the spin precession.
This geometric motion near the gap minima increases the effective gap $\Delta+\frac{FR_{12}}{2}$ and prevents tunneling.
More generally, the observed nonreciprocity is analogous to the well-known selective absorption of circularly polarized light, but in the non-perturbative regime.

As described above, the spin flips during the perfect tunneling. In terms of quantum control, a spin flip can also be achieved differently  using the Rabi oscillation and the adiabatic control (or its shortcut~\cite{guery2019shortcuts}). The driving field and spin are orthogonal, parallel, and antiparallel in the Rabi oscillation, the adiabatic control, and the TLZ model, respectively. 
This difference in the restriction of the driving field to the spin direction makes a difference in control speed, robustness, and implementability. Our geometric diabatic control is an effective means of increasing the versatility of quantum control (see~\cite{sm}).

\begin{figure}
\begin{center}
\includegraphics{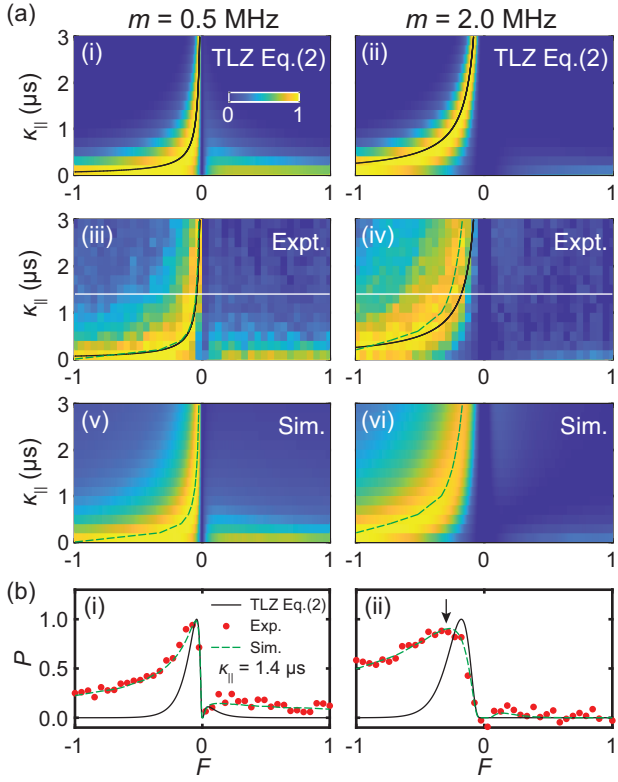}
\caption{
Gap parameter and curvature dependence of the TLZ transition probability.
(a) The left (right) panels denote the results at $m=0.5$~MHz ($m=2.0$~MHz).
The black solid (green dashed) line indicates the PT condition in the TLZ formula (simulation).
(b) Tunneling probability at $\kappa_{\parallel}=1.4~\mu$s [white line in (a iii) and (a iv)].
The black arrow in (ii) indicates the PT condition.
\label{fig3}
}\end{center}
\end{figure}

Next we study the validity of the TLZ formula [Eq.~(\ref{eq2})] when the twist becomes stronger; the higher-order terms ignored in the derivation of the TLZ formula increase and the precession is no longer perfectly synchronized with the quadratic twist.
We investigate the tunneling probability obtained at $m=0.5$~MHz for a curvature range from $\kappa_{\parallel}=0~\mu$s to $\kappa_{\parallel}=3~\mu$s.
Figure.~\ref{fig3}(a iii) shows the experimental result, representing a clear nonreciprocal behavior to the speed $F$.
As $\kappa_{\parallel}$ increases, the PT condition approaches zero.
A similar trend is observed in the TLZ formula shown in Fig.~\ref{fig3}(a i), indicating that this characteristic is consistent with $F_\text{PT} = -\frac{2\Delta}{R_{12}(0)}$.
This result proves that the speed of the quantum control is tunable by the geodesic curvature $\kappa_{\parallel}$ of the driving field.

For a more quantitative comparison, we show a cross section at $\kappa_{\parallel}=1.4~\mu$s in Fig.~\ref{fig3}(b i) [white line in Fig.~\ref{fig3}(a iii)].
The experimental result (red circles) exhibits $P\sim1$ near $F_\text{PT} = -0.045$ in good agreement with the TLZ formula (black solid line).
On the other hand, in (negatively) large speeds $F < F_\text{PT}$, $P$ decreases almost exponentially in the TLZ formula~\cite{takayoshi2021nonadiabatic}, whereas the change is gradual in the experimental result.
This deviation becomes prominent as the gap parameter $m$ and/or the curvature $\kappa_{\parallel}$ are larger.
The right panels of Figs.~\ref{fig3}(a) and (b) show the corresponding data sets obtained with a larger gap parameter ($m=2.0$~MHz).
The PT condition in the experiment (red circles) shifts to the left from what the TLZ formula (black solid line) predicts [black arrow in Fig.~\ref{fig3}(b ii)].
The maximum $P$ is then slightly suppressed from unity.

We obtain exact solutions by numerical simulations (see~\cite{sm}) to discuss this deviation.
The simulation results are in Figs.~\ref{fig3}(a v) and (a vi) and the green dashed lines in Fig.~\ref{fig3}(b).
They reproduce the experimental results satisfactorily over the entire speed range.
The black solid and green dashed lines in Fig.~\ref{fig3}(a) show the perfect tunneling conditions obtained by the TLZ formula and the simulation, respectively.
The results show that as the gap parameter $m$ and curvature $\kappa_{\parallel}$ become larger, the exact PT condition shifts toward the (negative) high speed side.
Our precise measurements reveal that the higher-order terms are essential for a quantitative understanding of the TLZ transition.

As shown above, we find that nonreciprocity and high tunneling probability at finite speed always persist even when the TLZ formula is invalid.
Thus, we conclude that these geometric effects are robust.
Introducing a field twist can be a ubiquitous method of adjusting tunneling probabilities at arbitrary speeds, making the present TLZ model an alternative framework for quantum control at various energy scales.
When applied to quantum materials, such control induces nontrivial properties such as the nonreciprocity of dc current and photocurrent \cite{kitamura2020nonreciprocal,takayoshi2021nonadiabatic}.

\begin{figure}
\begin{center}
\includegraphics{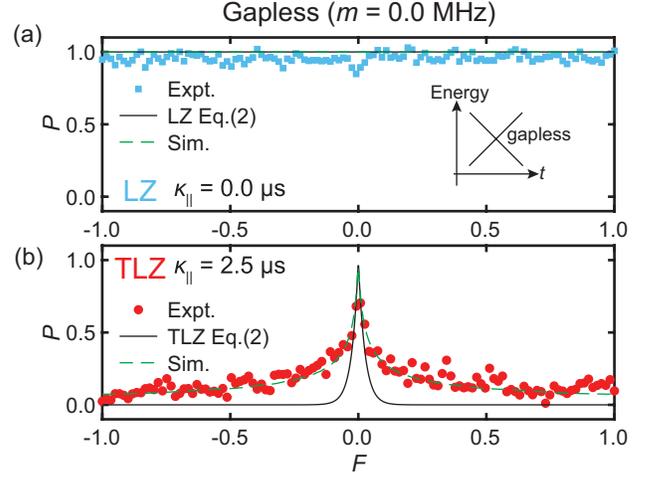}
\caption{
Sweep speed dependence of the transition probability of the gapless ($m=0.0$~MHz) system.
(a) The LZ transition ($\kappa_{\parallel}=0$).
The inset is a schematic of the energy change.
(b) The TLZ transition ($\kappa_{\parallel}=2.5~\mu$s).
\label{fig4}
}\end{center}
\end{figure}

In the case of an infinitesimal gap ($m=0.0$~MHz), 
the TLZ formula predicts a counter intuitive behavior, i.e., tunneling is suppressed as we increase the speed. 
Since this is relevant to the study of laser-field-driven dynamics in Dirac and Weyl semimetals~\cite{takayoshi2021nonadiabatic}, we study this situation in detail. 
The energy change is shown in the inset of Fig.~\ref{fig4}(a), which mimics the situation where electrons in the valence band accelerated by the electric field are excited through the Dirac (Weyl) point into the conduction band.
Here the LZ model and the TLZ model correspond to the case where the driving fields are dc and ac electric fields, respectively.
We examine the LZ model and observe that it yields $P\sim1$, as shown in Fig.~\ref{fig4} (a).
This is a straightforward phenomenon caused by the complete reversal of the field in the $y$ axis.
We then examine the TLZ transition at $\kappa_{\parallel}=2.5~\mu$s as in Fig.~\ref{fig4}(b).
The high tunneling probability near the adiabatic limit $F\sim0$ is consistent with $F_\text{PT}=-\frac{4\pi m}{v \kappa_{\parallel}}=0$ (for $m=0$).
This behavior, where the probability decreases with increasing sweep speed, is opposite to the LZ transition at a finite gap [Fig.~\ref{fig2}(b)].
This counter intuitive result is caused by the monocyclic nature of the quadratic twist, where the initial and final fields point in the same direction.
It is qualitatively reproduced by the TLZ formula (black solid line) and is perfectly reproduced in the simulation (green dashed line).

We experimentally confirmed the nonadiabatic geometric effects of nonreciprocity and perfect tunneling in the quadratic TLZ model over a wide range of parameters.
Specifically, we showed that we could utilize the geometric effects to control the quantum state dynamically. 
Geometric diabatic control can be applied to control systems of various energy scales, from nuclear spins to quantum materials.
An important challenge to improving this method is to 
find a way to enhance the tunneling probability and bring it even closer to 100~\%. 
We think this is possible by engineering the shape of the field twist to cancel the higher-order terms ignored in the derivation of the TLZ formula.

We thank K. M. Itoh (Keio University) for letting us use the confocal microscope system.
This work is partially supported by “Advanced Research Infrastructure for Materials and Nanotechnology in Japan(ARIM)” of the Ministry of Education, Culture, Sports, Science and Technorogy (MEXT), Proposal Number JPMXP1222UT1131.
This work is supported by JSPS Grants-in-Aid for Scientific Research (Nos.~JP22K03524, JP19H00656, JP19H05826, JP23H01103, JP20H02187, and JP20H05661), JST CREST (JPMJCR19T3 and JPMJCR1773), MEXT Q-LEAP (JPMXS0118068379),  JST Moonshot R\&D (JPMJMS2062), MIC R\&D for construction of a global quantum cryptography network (JPMI00316), Next Generation Artificial Intelligence Research Center at the University of Tokyo.

\nocite{jacques2009dynamic, dreau2011avoiding}


\begin{thebibliography}{39}
\bibitem [{sm()}]{sm}%
  \BibitemOpen
  \href@noop {} {\bibinfo {title} {See {S}upplemental {M}aterial for details,
  which includes {R}efs.~[38,39]}}\BibitemShut {NoStop}%
\end{thebibliography}

\providecommand{\noopsort}[1]{}\providecommand{\singleletter}[1]{#1}%

\cleardoublepage

\onecolumngrid

\renewcommand{\thefigure}{S\arabic{figure}}
\renewcommand{\theequation}{S\arabic{equation}}
\renewcommand{\thetable}{S\arabic{table}}
\setcounter{figure}{0}
\setcounter{equation}{0}

\section{Supplemental Material}

\subsection{TLZ Hamiltonian for the present experiment}

The expressions for the tunneling probability [Eq.~(2)] and PT condition [Eq.~(3)] obtained from the TLZ Hamiltonian [Eq.~(1)]  in the main text are derived in the natural unit ($\hbar = 1, c= 1$).
However, the MKS unit is more intuitive for actual experimentation.
A unit change in energy or time would require distorted parameter settings to take over the same parameters as in Eq.~(1).
Therefore, we use the following Hamiltonian with the Pauli matrix replaced by $S=1/2$ operators without changing the parameters of the driving field.
\begin{equation}
2\pi \hat{H}_\text{s} = 2\pi \bm{b}\cdot \bm{\hat{S}},\;\bm{b}=\qty(b_x,b_y,b_z)=\qty(m, \nu q, \frac{1}{2}\kappa_{\parallel} \nu^2q^2),
\label{eqms1}
\end{equation}
where, the unit of $m$ is Hz, the unit of $\nu$ is Hz$^2$, and the unit of $\kappa_{\parallel}$ is sec.

This corresponds to changing each parameter in Eq.~(1) in the main text as follows,
\begin{eqnarray}
m &\rightarrow& \pi m,  \nonumber \\
\nu &\rightarrow& \pi \nu,  \nonumber \\
\kappa_{\parallel} &\rightarrow& \kappa_{\parallel}/\pi.
\label{eqms2}
\end{eqnarray}
This changes the formula of the tunneling probability [Eq.~(2)] and PT condition [Eq.~(3)] as,
\begin{eqnarray}
P &\approx& \exp\left[-\frac{\pi}{4\nu \abs{F}}\left(\Delta+\frac{FR_{12}(0)}{2}\right)^2\right] 
\rightarrow
\exp\left[-\frac{\pi^2}{\nu \abs{F}}\left(m+\frac{F \nu \kappa_{\parallel}}{4\pi}\right)^2\right],
\label{eqms4}  \\
F_\text{PT} &=& -2\Delta/R_{12}(0) \rightarrow -\frac{ 4\pi m }{\nu \kappa_{\parallel}}.
\label{eqms5}
\end{eqnarray}
where $R_{12}(0)=\nu\kappa_{\parallel}$ is unchanged.
All experimental results (Figs.~2-4 in the main text) are analyzed with these formula.

\subsection{Sample and setup}
We utilize a $^{12}$C-enriched diamond (001) single crystal grown by chemical vapor deposition.
The dimension of the diamond sample is $3\times3\times0.4$~mm$^3$. 
The diamond contains nitrogen with a concentration of $20\sim40$~ppb, and the concentration of the NV center is inferred to be $1/100$ of the nitrogen concentration. 
The NV center showed a long quantum coherence $T_2 = 350~\mu$s with a resonance linewidth of 148~kHz.
By applying a static magnetic field of $B_0\sim50$~mT parallel to the NV symmetry axis ($+z$ direction), we polarize the nitrogen nuclear spin of the NV center~[38] and purely manipulate a two-level system consisting of $m_S=0$ and $m_S = -1$.
We perform the time-resolved PL measurements using laser confocal microscopy for spin initialization and readout.
The microwave waveforms are generated by IQ modulation with a vector signal generator (Tektronix TSG4104A) and an arbitrary waveform generator (Tektronix AWG7122C) to adjust amplitude, phase, and frequency.
A broadband microwave amplifier (Mini-Circuits ZHL-5W-63-S+) and a copper wire coplanar waveguide antenna are used to reduce the frequency dependence of the microwave.
We absorb the microwaves reflected by the coplanar waveguide with a circulator and a termination not to degrade the amplifier's characteristics.

\subsection{Microwave parameters}

We derive the Hamiltonian in a rotating frame [Eq.~(4) in the main text] based on the Hamiltonian of the NV center in the laboratory coordinate.
The Hamiltonian for the subspace of $m_S=0$ and $m_S=-1$ is expressed as,
\begin{eqnarray}
\hat{H}_\text{NV} &=& (-D+\gamma_\text{e}B_0)\hat{S}_z 
+ f_\text{R} \cos{\qty(2\pi f_\text{mw} t + \phi_\text{mw})} \hat{S}_x 
- f_\text{R} \sin{\qty(2\pi f_\text{mw} t + \phi_\text{mw})} \hat{S}_y,
\label{eqm1}
\end{eqnarray}
where $D=2.87$~GHz is the zero-field splitting, $\gamma_\text{e}=28$~GHz/T is the gyromagnetic ratio, $B_0$ is the static magnetic field strength, $f_\text{R}$ is the Rabi frequency corresponding to the microwave field strength, $f_\text{mw}$ is the microwave frequency, and $\phi_\mathrm{mw}$ is the microwave phase.
Since our experiment utilizes a static magnetic field strength $B_0\sim50$~mT, the total field $(-D+\gamma_\text{e}B_0)$ in the NV symmetry axis ($z$ direction) is negative, the spin precesses clockwise about the $z$ axis, and the resonance frequency is $f_\text{res}=D-\gamma_\text{e}B_0$.
Because we use a weak microwave ($f_\text{R} \ll f_\text{res}$), we applied a rotating wave approximation that considers only the clockwise circularly polarized component of the microwaves.

In a rotating frame around the $z$ axis, the effective field in the $z$ direction can be modulated because the spin precession speed increases or decreases with the rotational speed of the coordinate.
Specifically, in the rotating frame, the Hamiltonian can be rewritten as,
\begin{align}
\hat{H}_\text{r} &= \hat{U}^{\dagger} \hat{H}_\text{NV} \hat{U} - \frac{i}{2\pi}\hat{U}^{\dagger}\qty(\frac{d}{dt}\hat{U}),
\label{eqm2}
\end{align}
where $\hat{U}=\exp[ 2\pi i f_\text{mw}t \hat{S}_z ]$ is the operator rotating clockwise about the $z$ axis at microwave frequency.
The driving field in this coordinate system can be expressed as,
\begin{align}
\bm{b} &= \qty[ f_\text{R} \cos(\phi_\text{mw}), - f_\text{R} \sin(\phi_\text{mw}), \frac{d (f_\text{det}t)}{dt} ],
\label{eqm3}
\end{align}
where $f_\text{det}$ is the microwave detuning from the resonance frequency and satisfies $f_\text{mw}=f_\text{ref}+f_\text{det}$.
We set the microwave parameters appropriately to represent the driving field of the TLZ Hamiltonian [Eq.~(1) in the main text].

Although the sweep duration $T$ is assumed to be infinitely long in the derivation of the tunneling probability [Eq.~(2) in the main text], experimentally its maximum value is limited by available microwave parameter ranges and quantum coherence.
The maximum sweep duration that Eq.~(\ref{eqm2}) can reproduce the TLZ Hamiltonian is expressed as,
\begin{align}
T_\mathrm{R} &= 2\frac{ f_\mathrm{R}^\mathrm{max}}{\nu\abs{F}}, \label{eq9} \\
T_\mathrm{det} &= 2 \frac{ \sqrt{ 2 f_\mathrm{det}^\mathrm{max}/\abs{\kappa_{\parallel}} } }{\nu\abs{F}}, \label{eq10}
\end{align}
where $f_\text{R}^\text{max}$ and $f_\text{det}^\text{max}$ are the maximum available Rabi frequency and detuning, respectively.
In our experiments, we set these values sufficiently smaller than the Zeeman splitting under 50~mT ($f_\text{R}^\text{max}=13.6$~MHz and $f_\text{det}^\text{max}=50$~MHz) to ignore the transition between $m_S=0,+1$ states.
The maximum duration in the present experiment is set to $T=10~\mu$s, which is sufficiently shorter than the present $T_2$, and the shortest duration is adopted after comparing with $T_\text{R}$ and $T_\text{det}$.
Our simulations confirmed that the sweep duration is regarded as sufficiently long to reproduce Eq.~(2) in the main text.

We also describe the amplitude and phase settings of the rectangular pulses used in the initialization and readout.
In contrast to the rotating frame at the resonance frequency where the spin precession stops, the rotating frame for Eq.~(4) in the main text is faster by the amount of detuning $f_\text{det}$.
The rotation angle $\theta$ and phase $\phi$ to prepare the initial and readout states are expressed as,
\begin{align}
\theta_i &= \acos( b_z(0)/\abs{ \bm{b}(0) } ), \\
\phi_i &= -\qty[\frac{\pi}{2} + \atan(b_y(0)/b_x)], \\
\theta_f &= \acos(b_z(T)/\abs{\bm{b}(T)}), \\
\phi_f &= 2\pi f_\mathrm{det}(T) T - \qty[-\frac{\pi}{2} + \atan(b_y(T)/b_x)].
\end{align}
These microwave phases are defined in a rotating frame at the resonance frequency.
We control the spin rotation angle by adjusting the amplitude of the rectangular microwave pulses with a constant duration of 130~ns.

\subsection{Numerical simulations}

We use the ordinary differential equation solver of MATLAB\textsuperscript{\textregistered} to compute the spin state following the TLZ Hamiltonian with the time-dependent Schr\"odinger equation.
Our simulation assumes an $S=1/2$ system in a specific rotating frame [see Eq.~(\ref{eqm2})].
This situation assumes that a transition between $m_S = 0$ state and $m_S=-1$ state can be frequency-selectively controlled under an external magnetic field (see SAMPLE AND SETUP).
We straightforwardly compute the time evolution of the two-level system.
We include a finite sweep duration $T$ in the simulation, and find its effect is negligible except when speed $F\sim0$ (see next section).

\subsection{Effect of decoherence}

\begin{figure}
\begin{center}
\includegraphics{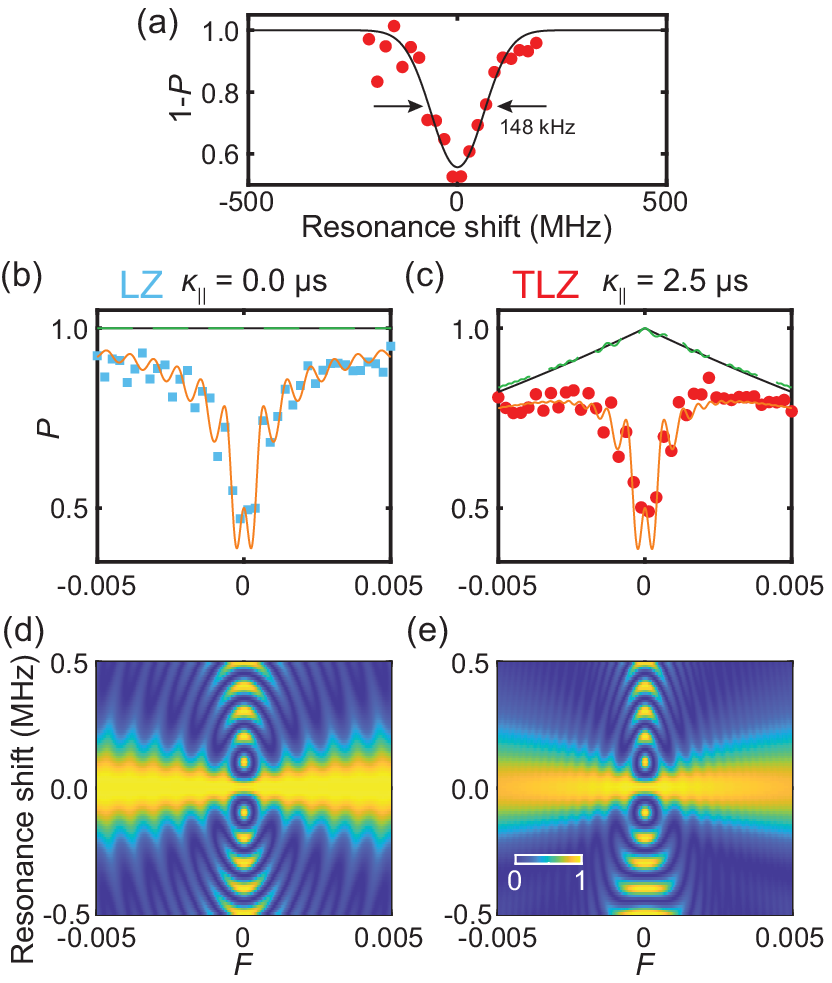}
\caption{
Effects of decoherence.
(a) The resonance shape of the current NV center measured by pulsed ODMR spectroscopy~[39] using a pulse of $T_{\pi}=28~\mu$s duration.
Red dot: experiment, solid black: Gaussian fit.
(b,c) Behavior near the adiabatic limit $F\sim0$ for LZ (b) and TLZ (c) transitions.
(d,e) Simulation of dependence on resonance shifts for LZ (b) and TLZ (c) transitions.
In (b,c,d,e), red and blue dots are the experimental data, solid black lines are the TLZ formula , dashed green lines are simulation, and solid orange lines are simulation that considers the decoherence.
\label{figs1}
}\end{center}
\end{figure}

Decoherence is an essential factor in the experiment that is not considered in the TLZ formula [Eq.~(2) in the main text] and the simulations presented in the main text.
This decoherence is caused by a stochastic resonance frequency shift of the electron spin due to interaction with the surrounding environment.
Thus, it appears as the linewidth of the resonance spectrum.
Figure~\ref{figs1}(a) shows the resonance spectrum of the present NV center.
The experimental result (red dot) is well reproduced by the Gaussian (solid black line) with a full width at a half maximum (FWHM) of 148~kHz.
In the coordinate system used in this study, this noise produces an offset in the $z$ component of the driving field.

This effect becomes vital near the adiabatic limit $F\sim0$ of the gapless systems.
Figure~\ref{figs1}(b) shows the LZ transition of the gapless system, obtained at a higher resolution than those in Fig.~4 in the main text.
The experimental data (blue dot) show an oscillatory suppression of the transition probability to 50~\% as the speed decreases ($F\rightarrow0$).
This behavior is qualitatively different from the TLZ formula  (solid black line) and simulation (dashed green line).

We find that the decoherence causes this suppression.
Figure~\ref{figs1}(d) is a colormap of the dependence on the resonance shift of the two-level system.
As the absolute shift increases, the probability decreases because the gap increases.
We numerically simulate the influence of decoherence by weighting and integrating the simulation results with a normalized resonance shape [Fig.~\ref{figs1}(a)].
Specifically, we calculated the transition probability as,
\begin{align}
P_\text{lw} &= \int \rho(f_\text{s}) P(f_\text{s}) d f_\text{s},
\end{align}
where $\rho(f_\text{s})$ is the normalized Gaussian with FWHM of 148~kHz and $P(f_\text{s})$ is the tunneling probability obtained in the simulation at resonance shift of $f_\text{s}$.
The result is shown as the solid orange line in Fig.\ref{figs1}(b), which is in quantitative agreement with the experimental result.
This explains the decoherence effect during the transition, which is essentially different from the decoherence in the LZ interference~(for example Ref.~[12]).
Our simulations assume that the noise frequency is sufficiently low and that the resonance shift does not change during a single driving field sweep.
This assumption is consistent with the fact that a long $T_2$ is obtained since such low-frequency noise is canceled out by the refocusing in the Hahn echo sequence used to measure $T_2$.

The corresponding data sets for the TLZ transition at $\kappa_{\parallel}=2.5~\mu$s are shown in Fig.~\ref{figs1}(c) and Fig.~\ref{figs1}(e).
The perfect tunneling near the adiabatic limit $F = 0$ appears in the TLZ formula  (solid black line) and simulation (dashed green line), whereas it is suppressed in the experiment (red dot).

The noise effect we observed is crucial in the driving of quantum materials.
A realistic Dirac (Weyl) system must have a finite energy width at the Dirac (Weyl) point.
Corresponding to our experimental results, this energy width gives the lower frequency limit at which the electric field to drive the carriers will function.
Strictly, depending on the quantum system, noise may appear not only in the quadratically varying components of the driving field but also in the gap and linearly varying components.
The frequency of the noise may also vary.
These details lead to behavior that is quantitatively different from the present experimental results but beyond this study's scope.

\subsection{Effect of microwave imperfection}

\begin{figure}
\begin{center}
\includegraphics{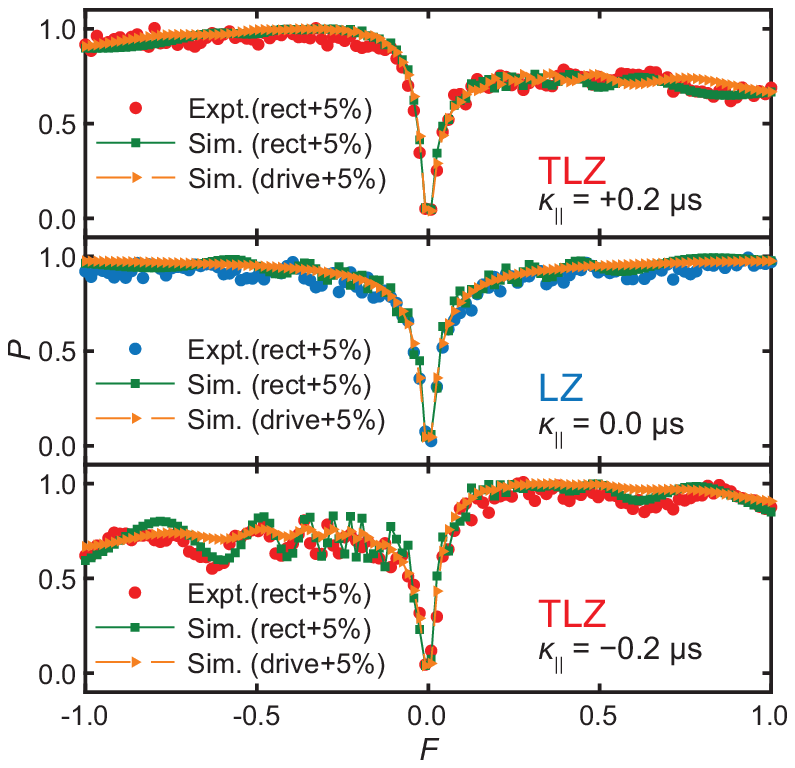}
\caption{
The effect of microwave amplitude error on tunneling probability.
The top, middle, and bottom panels are the results at curvatures $\kappa_{\parallel} = -0.2, 0$, and $0.2~\mu$s, respectively.
Red dot: experimental data for a 5\% increase in rectangular pulse amplitude; green square: simulation for a 5~\% increase in rectangular pulse amplitude; orange triangle: simulation for a 5~\% increase in pulse amplitude for the driving field.
\label{figs2}
}\end{center}
\end{figure}

The microwave circuits, such as the amplifier and the antenna, generally have nonlinear amplification and frequency dependence on amplitude.
This imperfection causes amplitude errors in microwave pulses.
The green square and orange triangle in Fig.~\ref{figs2} are the simulations where only the rectangular pulse and the chirp pulse representing the driving field have an amplitude error of +5~\%, respectively.
The top, middle, and bottom panels show the results at curvatures $\kappa_{\parallel} = -0.2, 0$, and $0.2~\mu$s, respectively.
The noise of the rectangular pulses modulates the tunneling probability to a greater extent than the chirp pulses.
The amplitude error of the rectangular pulses results in an error in the spin direction at the initial and final states.
We focus on the $F<0$ regions in the $\kappa_{\parallel} = -0.2~\mu$s case [lower panel of Fig.~\ref{figs2}], where the oscillations are particularly large.
This is the area where the rectangular pulses produce extensive rotations.
The red and blue dots in Fig.~\ref{figs2} are the experimental results when the rectangular pulse amplitude is 5~\% larger than in the experiments presented in the main text.
The modulation has a larger amplitude than that in Fig.~2(b) in the main text and exhibits qualitatively the same behavior as in the simulation (green squares).
In detail, the oscillation period of this modulation is somewhat slower with respect to the sweep speed than in the simulation.
An error in the microwave corresponding to the driving field near the gap minima causes such an effect.
This error also shifts the PT condition.
We confirm that the simulations with ideal pulses yield an average tunneling probability of 99.7~\% for the curvatures range of Fig.~2(c) in the main text.
The deviation from 100~\% in Fig.~2(c) in the main text is caused by this shift of the PT condition.

\subsection{Comparison with other control methods}

\begin{figure}
\begin{center}
\includegraphics{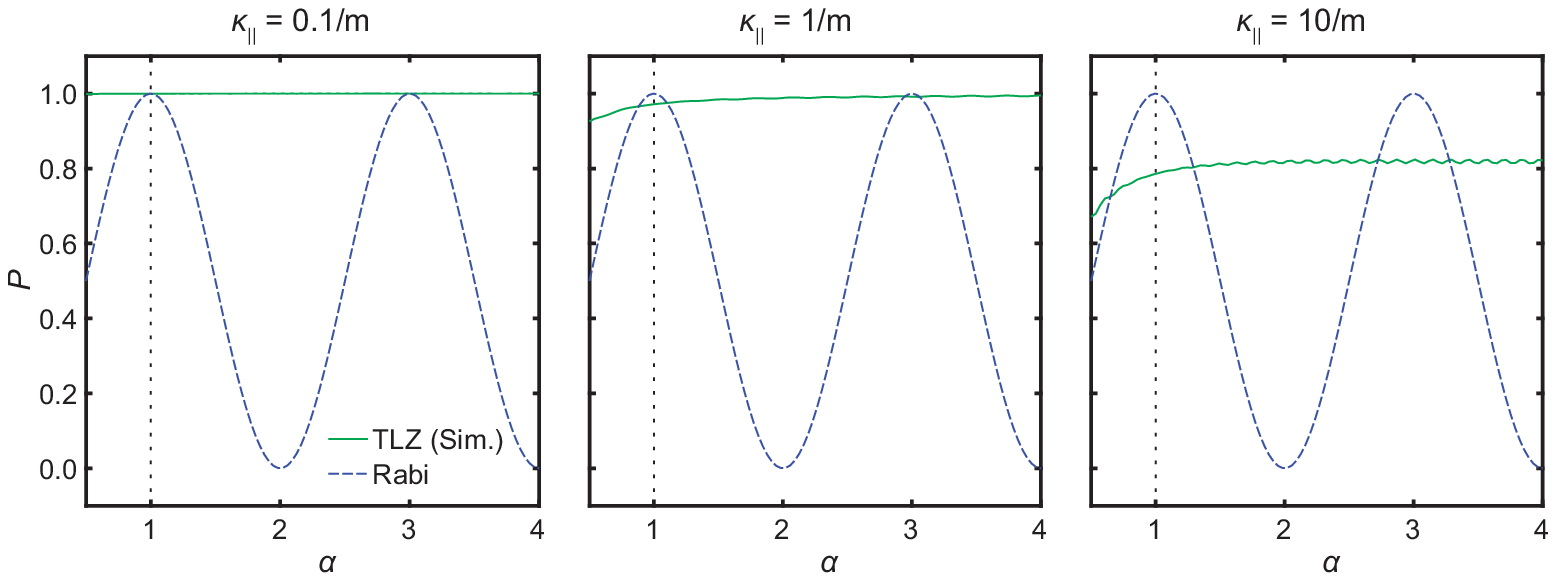}
\caption{
The driving amplitude error dependence of transition probability.
The left, center, and right panels are the results for curvatures of $\kappa_\parallel = 0.1/m$, $\kappa_\parallel = 1/m$, and $\kappa_\parallel = 10/m$, respectively.
The amplitude errors are more significant as $\alpha$ deviates from 1.
The sweep speed $F$ in our TLZ model is set such that the maximum transition probability is obtained in the error-free condition ($\alpha = 1$).
\label{figs3}
}\end{center}
\end{figure}

Here we compare our geometric non-adiabatic control (TLZ model), an adiabatic control, and a resonant control.

Our TLZ model and adiabatic controls differ in bringing the transition probability closer to 100~\% and 0~\%, respectively.
This difference can be critical in implementability.
For example, if we wants to excite carriers from the valence band to the conduction band, adiabatic control is difficult because it is usually impossible to drive band inversion.
On the other hand, in principle, our TLZ model can realize diabatic transitions across the band gap.

Our TLZ model differs from a resonant control in terms of robustness.
In the case of resonant Rabi oscillations, the transition probability is given by,
\begin{align}
P &= \frac{1}{2} [1-\cos{(2\pi f_\text{R} T )} ],
\end{align}
where $f_\text{R}$ is Rabi frequency, and $T$ is control duration.
A perfect transition can be implemented by setting the rotation angle $\theta = 2 \pi f_\text{R} T = \pi$.
Since the Rabi frequency $f_\text{R}$ varies linearly with the drive amplitude, the transition probability is degraded in the presence of the amplitude error.
To quantify the influence, we define an error factor $\alpha$ that changes $\theta=\pi \rightarrow \alpha \pi$.
Here, there is no error when $\alpha = 1$, and the more it deviates from 1, the larger the error.

Our TLZ model is robust against the amplitude error.
As with Rabi oscillations, we assume that the error appears in the driving amplitudes other than gap, ($b_y= \nu q \rightarrow \alpha \nu q$, $b_z = \frac{1}{2}\kappa_\parallel \nu^2 q^2 \rightarrow \alpha \frac{1}{2}\kappa_\parallel \nu^2 q^2$).
The change of driving fields can be considered as the following parameter changes; $\nu \rightarrow \alpha \nu$ and $\kappa_\parallel \rightarrow \kappa_\parallel/\alpha$.
Interestingly, it does not modify the perfect tunneling condition [Eq.~(\ref{eqms5})].
Thus, it should be robust against the amplitude error.

To quantitatively compare the robustness of our TLZ model with the Rabi oscillation, we show simulation results in Fig.~\ref{figs3}.
We examine the transition probability for a wide range of the amplitude error from 0.5 to 4.
The left and center panels show the results for $\kappa_\parallel = 0.1/m$ and $\kappa_\parallel = 1/m$, respectively.
The transition probability using the Rabi oscillation degrades steeply as $\alpha$ deviates from 1 and it oscillates with respect to $\alpha$. 
On the other hand, as expected, our TLZ model shows a wide error range that can keep the transition probability above 90~\%.
This means that our TLZ model is overwhelmingly robust against amplitude errors.

The right panel shows the results for $\kappa_\parallel = 10/m$.
Our TLZ model shows a wide error range that can keep the transition probability above 80~\%.
Comparing to other panels, the transition probability degrades with increasing $\kappa_\parallel$.
This is due to a breaking of the approximation in our TLZ model [see also Fig.~3 in the manuscript].
The slight increase in the transition probability when $\alpha$ is greater than 1 is due to the reduction in effective curvature, which mitigates the approximation breaking.
Increasing the curvature $\kappa_\parallel$ in our TLZ model leads to faster state transition.
Therefore, further understanding of approximation breaking in our TLZ model may allow us to design even faster, more robust, and higher precision controls.

\end{document}